\documentclass[12pt]{article}
\usepackage{geometry}
\usepackage{amsmath}
\usepackage{amssymb}
\usepackage{graphicx}
\numberwithin{equation}{section}

\def\appendix#1{
  \addtocounter{section}{1}
  \setcounter{equation}{0}
  \renewcommand{\thesection}{\Alph{section}}
 \section*{Appendix \thesection\protect\indent \parbox[t]{11.715cm} {#1}}
  \addcontentsline{toc}{section}{Appendix \thesection\ \ \ #1}
  }

\renewcommand{\thefootnote}{\fnsymbol{footnote}}

\newcommand{\be}{\begin{equation}}
\newcommand{\ee}{\end{equation}}
\newcommand{\ba}{\begin{aligned}}
\newcommand{\ea}{\end{aligned}}

\def\m1{\left(-1\right)^{F_i}}

\makeatletter
\def\sla@#1#2#3#4#5{{%
  \setbox\z@\hbox{$\m@th#4#5$}%
  \setbox\tw@\hbox{$\m@th#4#1$}%
  \dimen4\wd\ifdim\wd\z@<\wd\tw@\tw@\else\z@\fi
  \dimen@\ht\tw@
  \advance\dimen@-\dp\tw@
  \advance\dimen@-\ht\z@
  \advance\dimen@\dp\z@
  \divide\dimen@\tw@
  \advance\dimen@-#3\ht\tw@
  \advance\dimen@-#3\dp\tw@
  \dimen@ii#2\wd\z@  \raise-\dimen@\hbox to\dimen4{%
    \hss\kern\dimen@ii\box\tw@\kern-\dimen@ii\hss}%
  \llap{\hbox to\dimen4{\hss\box\z@\hss}}}}
\def\slashed#1{%
  \expandafter\ifx\csname sla@\string#1\endcsname\relax
    {\mathpalette{\sla@/00}{#1}}%
  \else
    \csname sla@\string#1\endcsname
  \fi}
\makeatother

\def\ccw{{\hspace{-2.5mm}\unitlength 0.1in
\begin{picture}(1.00,1.00)(8.45,-11.50)
\special{pn 8}%
\special{pa 1120 1120}%
\special{pa 1100 1100}%
\special{pa 1080 1120}%
\special{fp}%
\end{picture}%
\hspace{-0mm}}}

\def\cw{{\hspace{-2.5mm}\unitlength 0.1in
\begin{picture}(1.00,1.00)(8.45,-11.50)
\special{pn 8}%
\special{pa 1120 1100}%
\special{pa 1100 1120}%
\special{pa 1080 1100}%
\special{fp}%
\end{picture}%
\hspace{-0mm}}}

\newcommand{\beq}{\begin{equation}}
\newcommand{\eeq}{\end{equation}}
\newcommand\beqa{\begin{eqnarray}}
\newcommand\eeqa{\end{eqnarray}}
\newcommand\bea{\begin{array}}
\newcommand\eea{\end{array}}

\newcommand{\nn}{\nonumber}
\newcommand{\neqa}{\nonumber\end{eqnarray}}
\newcommand{\la}{\label}

\newcommand{\color}[1]{}

\newcommand{\h}{\hat}
\renewcommand{\t}{\tilde}

\def\({\left(}
\def\){\right)}
\def\[{\left[}
\def\]{\right]}

\def\<{\langle}
\def\>{\rangle}

\def\d{\partial}

\begin{document}


\thispagestyle{empty}
\begin{flushright}\footnotesize
\texttt{CALT-68-2667}\\
\texttt{LPTENS 08/04}\\
\texttt{SPhT-t08/017}\\
\vspace{2.1cm}
\end{flushright}

\renewcommand{\thefootnote}{\fnsymbol{footnote}}
\setcounter{footnote}{0}

\begin{center}
{\Large\textbf{\mathversion{bold} Quantum Wrapped Giant Magnon}\par}

\vspace{2.1cm}

\textrm{Nikolay Gromov$^{\alpha}$, Sakura Sch\"afer-Nameki$^{\beta}$ and Pedro Vieira$^{\gamma}$}
\vspace{1.2cm}

\textit{$^{\alpha}$ Service de Physique Th\'eorique,
CNRS-URA 2306 C.E.A.-Saclay, F-91191 Gif-sur-Yvette, France;
Laboratoire de Physique Th\'eorique de
l'Ecole Normale Sup\'erieure et l'Universit\'e Paris-VI,
Paris, 75231, France;
St.Petersburg INP, Gatchina, 188 300, St.Petersburg, Russia } \\
\texttt{nikgromov@gmail.com}
\vspace{3mm}

\textit{$^{\beta}$ California Institute of Technology\\
1200 E California Blvd., Pasadena, CA 91125, USA } \\
\texttt{ss299@theory.caltech.edu}
 \vspace{3mm}

\textit{$^{\gamma}$ Laboratoire de Physique Th\'eorique
de l'Ecole Normale Sup\'erieure et l'Universit\'e Paris-VI, Paris,
75231, France;  Departamento de F\'\i sica e Centro de F\'\i sica do
Porto Faculdade de Ci\^encias da Universidade do Porto Rua do Campo
Alegre, 687, \,4169-007 Porto, Portugal} \\
\texttt{pedrogvieira@gmail.com}
\vspace{3mm}


\par\vspace{1cm}

\textbf{Abstract}\vspace{5mm}
\end{center}

\noindent
Understanding the finite-size corrections to the fundamental excitations of a theory
is the first step towards completely solving for the spectrum in finite volume.
We compute the leading exponential correction to the quantum energy of the fundamental excitation of the light-cone gauged string in $AdS_5 \times S^5$, which is the giant magnon solution. We present two independent ways to obtain this correction: the first approach makes use of the algebraic curve description of the giant magnon. The second relies on the purely field-theoretical L\"uscher formulas, which depend on  the world-sheet S-matrix.
We demonstrate the agreement to all orders in $(\Delta/\sqrt{\lambda})^{-1}$ of these approaches, which in particular presents a further test of the S-matrix. We comment on generalizations of this method of computation to other string configurations.

\vspace*{\fill}

\setcounter{page}{1}
\renewcommand{\thefootnote}{\arabic{footnote}}
\setcounter{footnote}{0}

 \newpage



\newpage

\section{Introduction and Summary}

In the AdS/CFT correspondence we find ourselves at the point where the S-matrix  of \cite{Staudacher:2004tk, Beisert:2005fw,Beisert:2005tm, Beisert:2006ib, Arutyunov:2006yd} is believed to accurately describe the infinite-volume theory, albeit it fails to capture some finite-size effects \cite{Ambjorn:2005wa,  SchaferNameki:2006gk, SchaferNameki:2005is, Arutyunov:2006gs, Janik:2007wt}.
The first step towards understanding a theory in finite volume is to compute the leading correction to the dispersion relation of its fundamental excitations. These corrections arise through virtual particles circling in the compact direction, which contribute to the self-energy of physical particles \cite{Luscher:1985dn}.
In this paper we compute the leading quantum finite-size corrections to the fundamental excitation of the $AdS_5 \times S^5$ string theory.

Tree-level light-cone gauged string theory on $AdS_5 \times S^5$ is a two-dimensional integrable field theory defined on a worldsheet cylinder of length $L/\sqrt{\lambda}$, where $L$ is a large symmetry charge and $1/\sqrt{\lambda}$, with $\lambda$ the t'Hooft coupling, plays the role of $\hbar$. In the infinite length limit the fundamental excitations are worldsheet solitons, so-called giant magnons (GM) \cite{Hofman:2006xt}, which transform under the residual extended $SU(2|2)^2$ symmetry \cite{Beisert:2005tm} and have a non-relativistic dispersion relation\footnote{It can also be obtained from a relativistic theory by integrating out some physical degrees of freedom \cite{Gromov:2006dh}.}
\beq
\epsilon_\infty(p)=\sqrt{1+\frac{\lambda}{\pi^2} \sin^2\(\frac{p}{2}\)} \la{infvol} \,,
\eeq
where $p$ is the magnon worldsheet momentum. This infinite volume dispersion relation 
is believed to be exact and
almost fixed by symmetry  alone  \cite{Beisert:2005tm}. For large $\lambda$
\beq
\epsilon_\infty(p)=\frac{\sqrt{\lambda}}{\pi} \sin\(\frac{p}{2}\) +0+\mathcal{O}\(\frac{1}{\sqrt{\lambda}}\) \la{exp} \,,
\eeq
where the leading term is the classical energy of the giant magnon and the absence of the $\mathcal{O}(1)$ term means, that the first quantum correction (or one-loop shift) vanishes for the magnon in infinite volume. This was explicitly checked in \cite{Papathanasiou:2007gd,Chen:2007vs}.

The finite size corrected dispersion relation differs from the infinite volume one by exponentially suppressed terms which can be organized according to world-sheet loop order
\beq
\epsilon(p)-\epsilon_\infty(p)=\sqrt{\lambda} \,\delta \epsilon_{cl}+  \delta \epsilon_{1-loop}+\frac{1}{\sqrt{\lambda}} \delta \epsilon_{2-loop} + \dots \,. \nn
\eeq
The  classical finite size corrections to the dispersion relation were computed in\footnote{This was also preformed in a more controlled orbifold setup in \cite{Astolfi:2007uz} and generalized for bound states of magnons, called dyonic magnons \cite{Dorey:2006dq}, in \cite{Hatsuda:2008gd,Minahan:2008re}.} \cite{Arutyunov:2006gs} to be
\beq
\delta \epsilon_{cl}=-\frac{4}{\pi} \sin^3\frac{p}{2}e^{-\frac{2\pi \Delta}{ \sqrt{\lambda} \sin\frac{p}{2}}} +\mathcal{O}\(e^{-2\frac{2\pi \Delta}{ \sqrt{\lambda} \sin\frac{p}{2}}}\) \,,\la{cl}
\eeq
where $\Delta=L+\frac{\sqrt{\lambda}}{\pi} \sin\frac{p}{2}$.
The first quantum corrections to the dispersion relation (which will in this case be the leading term since the infinite volume contribution vanishes) are computed in this paper and are given by (\ref{CurveResult}), of which the first terms in the $(\Delta/\sqrt{\lambda})^{-1}$ expansion are\footnote{Notice
that (\ref{CurveResult}) is perfectly well-behaved for $p=\pi$. The
singularity in (\ref{qt}) at $p=\pi$ is due to the fact that the
saddle point approximation breaks down, because the log branch-points
approach the saddle point. The approximation (\ref{qt}) of
(\ref{CurveResult}) is valid as long as $|p - \pi| \gg e^{-
\pi\Delta/\sqrt{\lambda}}$.} 
\beqa
\delta \epsilon_{1-loop} =
 \frac{8  \sin^2\frac{p}{4}\,e^{-\frac{2\pi\Delta}{\sqrt{\lambda}}} }{\pi
   \left(\sin \frac{p}{2}-1\right)\({\Delta \over \sqrt{\lambda} }\)^{1/2} }\!\left[1
     \!  - 
   \frac{7\!+\! 4 \sin p-\!4\cos p\!+\sin\frac{p}{2}
  }{16\pi \left(\sin\frac{p}{2}-\!1\right){\Delta \over \sqrt{\lambda} }}
   +\mathcal{O}\!\left(\!{1\over\({\Delta \over \sqrt{\lambda} }\)^{2}}\!   \right)
\right]\! +\dots ,\la{qt}
\eeqa
where the subleading corrections are of order 
$$\mathcal{O}\!\(\!e^{-\frac{2\pi\Delta}{\sqrt{\lambda}\sin\frac{p}{2}}},e^{-\frac{4\pi\Delta}{\sqrt{\lambda}}}\!\)\,.$$
As will be explained in section \ref{sec:Curve} this computation relies solely on the algebraic curve  technology  \cite{Kazakov:2004qf} and can be easily generalized to other states.

\begin{figure}[t]
    \centering
        \resizebox{90mm}{!}{\includegraphics{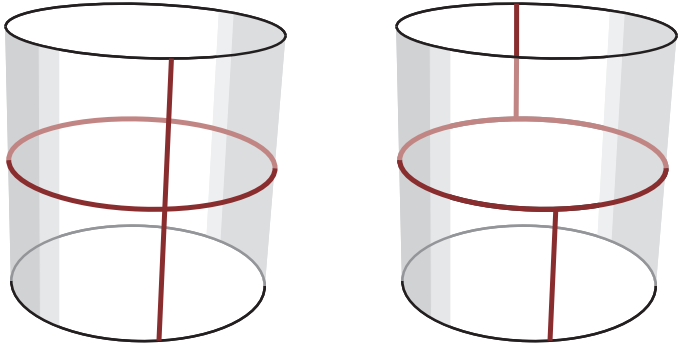}}
        \caption{\small The two leading processes contributing to the corrections of the dispersion relation in finite volume: The F-term (LHS) describes the contribution of a virtual particle loop. We shall see that it accounts for the leading contribution or  $\mathcal{O}\(e^{-2\pi \Delta/\sqrt{\lambda}}\)$ correction to the giant magnon one-loop energy. The $\mu$-term (RHS) is the effect of a particle splitting into two on-shell particles and computes the classical finite-size effects. It also contributes to the $1$-loop shift, but since it is $\mathcal{O}\(e^{-2\pi \Delta/\sqrt{\lambda}\sin\frac{p}{2}}\)$, this contribution is subleading.
        \label{figFmu}}
\end{figure}

On the other hand it is also known that the leading corrections to the dispersion relation of particles in finite volume is given by L\"uscher terms and are a sum of an F-term and a $\mu$-term contribution depicted in figure \ref{figFmu}, which are expressed in terms of the two particle $S$-matrix. The match with the corrections (\ref{cl}) and (\ref{qt}) provides therefore an important non-trivial check of the AdS/CFT S-matrix
\cite{Staudacher:2004tk, Beisert:2005fw,Beisert:2005tm, Beisert:2006ib, Arutyunov:2006yd}. The leading classical correction (\ref{cl}) was shown to match the $\mu$-term computation \cite{Janik:2007wt, Hatsuda:2008gd,Minahan:2008re} and we shall show that the leading quantum exponential correction (\ref{qt}) is accounted for by the F-term contribution \cite{Luscher:1985dn, Janik:2007wt}\footnote{Note that this formula differs from the one in \cite{Janik:2007wt} by the supertrace, i.e. insertion of $(-1)^F$ coming from the fermion loops. We will discuss this point in detail in appendix B.}
\beq\label{FtermdeltaE}
\delta E_a^F= - \int_{\mathbb{R}} \frac{dq}{2\pi} \(1-\frac{\epsilon_{\infty}'(p)}{\epsilon_{\infty}'(q^*(q))}\)e^{-iq^*(q) L}\sum_{b} \,(-1)^{F_b}\,(S_{ba}^{ba}(q^*(q),p) - 1) \,,
\eeq
where $p$ and $q$ are the momenta of the magnon that correspond to the physical particle and to the virtual particle propagating in the loop, respectively, and $q^*$ is determined by the onshell condition
$q^2+\epsilon_{\infty}^2(q_*(q))=0$.

In section \ref{sec:Fterm} we check that the F-term contribution yields the quantum result to all orders in $(L/\sqrt{\lambda})^{-1}$, which agrees with the algebraic curve result that we obtain in section 2!


\section{One-loop energy shift from the algebraic curve}\label{sec:Curve}

\subsection{GM from the algebraic curve}

The algebraic curve formalism \cite{Kazakov:2004qf} maps classical string solutions in $AdS_5\times S^5$ to a set of eight quasimomenta $\{p_{\t 1},\dots, p_{\t 4},  p_{\h 1},\dots,  p_{\h 4}\}$ which define an eight-sheeted Riemann surface.
It is convenient to embed the simple giant magnon solution into a family of solutions, called dyonic giant magnons \cite{Dorey:2006dq}, whose infinite volume dispersion relation is given by
$$
\epsilon_\infty^Q(p)=\sqrt{Q^2+16 g^2 \sin^2\(\frac{p}{2}\)}\,,
$$
where the momentum $p$ of the magnon should obviously not be confused with the quasimomenta $p_i$. For $Q=1$ the dyonic magnon reduces to the simple magnon mentioned in the introduction.

\begin{figure}[t]
    \centering
        \resizebox{70mm}{!}{\includegraphics{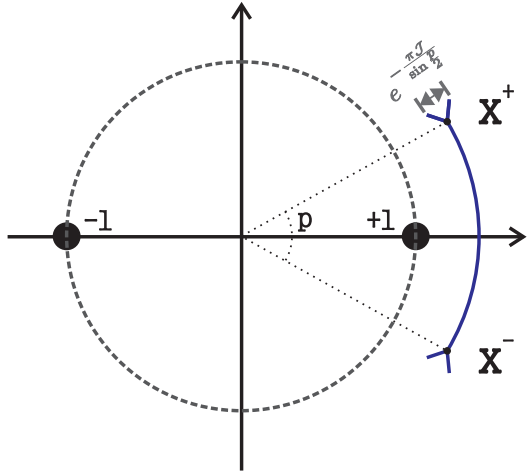}}
        \caption{\small The small charge dyonic giant magnon in finite volume \cite{Vicedo:2007rp}: the endpoints of the log-cut between $X^+$ and $X^-$ develop small square root tails, separated by $\mathcal{O}\(e^{- {\pi \mathcal{J} \over \sin p/2}}\)$, which will induce finite size corrections of this order. For the classical finite-size corrections, these were the only contributions. However, as we explain in the main text, for the leading quantum corrrection there are more important corrections, which are of order $\mathcal{O}\(e^{-2\pi \mathcal{J}}\)$.
                \la{figCurve}}
\end{figure}

In terms of the variables $X^\pm$ which parametrize the magnon momentum as
\beq
p=\frac{1}{i}\log\frac{X^+}{X^-} \,, \la{momQ}
\eeq
and are constrained by
\beq
Q=\frac{\sqrt{\lambda}}{4\pi i} \(X^++\frac{1}{X^+}-X^--\frac{1}{X^-}\) \,, \la{QQ}
\eeq
the infinite volume dispersion relation takes the form
\beq
\epsilon_\infty^{Q}= \frac{\sqrt{\lambda}}{4\pi i} \(X^+ - \frac{1}{X^+}- X^-+ \frac{1}{X^-} \) \,. \la{epsQ}
\eeq
The $X^\pm$ variables are particularly suitable for the curve treatment described below.
Notice that for a single magnon, or more
generally for $Q\ll \sqrt{\lambda}$ which is the case we consider in
this paper,
\beq\label{Xofp}
X^{\pm} = e^{\pm i p/2}  + \mathcal{O}(1/\sqrt{\lambda}) \,,
\eeq
and thus $X^\pm=1/X^\mp$.

The precise description \cite{Vicedo:2007rp} of the dyonic giant magnon in finite volume is that of a two-cut solution, where the two cuts are very close and thus can be reconnected to form a log-cut condensate with log branch points at $X^\pm$, see figure \ref{figCurve}. The separation of the two branch-points of the two-cut solution is of the order\footnote{This result holds for $Q\ll \sqrt{\lambda}$. The generalization for bigger $Q$ is also know \cite{Minahan:2008re}.}
$ \mathcal{O}\(e^{- \pi \Delta/ \sqrt{\lambda} \sin\frac{p}{2}} \)$.
The classical energy of the giant magnon in finite volume differs from the infinite volume log result (\ref{epsQ}) by a contribution of the order 
 \cite{Arutyunov:2006gs,Minahan:2008re},
$$  \mathcal{O}\( e^{-\frac{2\pi \Delta}{ \sqrt{\lambda} \sin\frac{p}{2}}} \) \,. $$
In the same way, if we take into account the precise two-cut structure of the finite volume GM we will also find a contribution to the fluctuation energies -- and therefore to the one-loop shift -- with the same exponential suppression. However, as already anticipated in the introduction, there are more important corrections (\ref{qt}) of the order
\beq
\mathcal{O}\(e^{-\frac{2\pi \Delta}{\sqrt{\lambda}} } \) \,, \nn
\eeq
which simply come from the fact that in order to compute the ground state energy
 in finite volume,  we should sum over the fluctuation energy mode number instead of integrating over the momenta of the quantum fluctuations. Thus, to compute the leading order quantum finite size corrections we can work with the log cut description
 \cite{Minahan:2006bd} of the giant magnon solution for which \cite{Minahan:2006bd, Chen:2007vs}
\beqa
\nn p_{\h 1,\h2}(x)=- p_{\h 3,\h4}(x)
&=&  {2 \pi \Delta \over \sqrt{\lambda}} \frac{x}{x^2-1}  \\
\nn
p_{\t 2}(x)\,\,\,=\,\,\,- p_{\t 3}(x)&=&{2 \pi \Delta \over \sqrt{\lambda}}
{ x\over x^2-1} +\frac{1}{i}  \log\frac{x-X^+}{x-X^-} \,\,\,\,\,\,+\t\phi_2\\
\la{ps}  
p_{\t 1}(x)\,\,\,=\,\,\,- p_{\t 4}(x)&=& {2 \pi \Delta \over \sqrt{\lambda}} \frac{x}{x^2-1} + \frac{1}{i}   \log\frac{x-1/X^-}{x-1/X^+}+   \t\phi_1 \,.
\eeqa
The twists $\t \phi_{1,2}$ in the quasimomenta \cite{Gromov:2007ky} are necessary because for a single magnon we cannot satisfy the usual periodic boundary conditions.  The usual orbifold treatment \cite{Ideguchi:2004wm,Beisert:2005he,Astolfi:2007uz} selects the twists
\beq
\t \phi_1=\t \phi_2=-p/2 \la{twists}\,,
\eeq
as reviewed in appendix $C$.

We shall now study the quantum fluctuations around this classical solution.
By perturbing the quasimomenta via the addition of extra poles we can compute the fluctuation energies $\Omega_n^{ij}$ around this classical solution \cite{Gromov:2007aq}. The poles are shared by quasimomenta $p_i$ and $p_j$ and the different possible choices of quasimomenta correspond to different string polarizations, namely
\beqa\label{Stacks}
S^5:       && (\tilde{1}\tilde{3})\,,\ (\tilde{1}\tilde{4})\,,\ (\tilde{2}\tilde{3})\,,\ (\tilde{2}\tilde{4})\,,\cr
AdS^5:   && (\hat{1}\hat{3}) \,,\ (\hat{1}\hat{4}) \,,\ (\hat{2}\hat{3}) \,,\ (\hat{2}\hat{4}) \,, \cr
\hbox{Fermions}:  && (\hat{1}\tilde{3}) \,,\  (\hat{1} \tilde{4}) \,,\ (\hat{2}\tilde{3}) \,,\ (\hat{2}\tilde{4}) \,,\cr
                 && (\tilde{1}\hat{3}) \,, \ (\tilde{1}\hat{4}) \,, \ (\tilde{2}\hat{3}) \,, \ (\tilde{2}\hat{4}) \,.
\eeqa
Moreover each quantum fluctuation has a mode number $n$, which will dictate where the corresponding pole outside the unit circle will be located through
\beq
p_i(x_n)-p_j(x_n)=2\pi n \,. \la{map}
\eeq
In this way we can find the semi-classical spectrum of fluctuations around the GM solution which reads
\beq \label{deltaDelta}
\delta \Delta= \sum_{(ij),n} N_n^{ij} \Omega_n^{ij} \,.
\eeq
The result of the perturbation of the quasimomenta, which is explicitly presented in appendix $B$, is that all fluctuation energies are given by the same function, when expressed in terms of the position of the poles $x_n$, that is $\Omega_n^{ij}=\Omega(x_x^{ij})$ for all $ij$. The function $\Omega(x)$ is given by
\beq\label{Om}
\Omega(x)=\frac{2}{x^2-1}\(1-\frac{X^++X^-}{X^+X^-+1}\, x\)\,.
\eeq
Notice that this does \textit{not} mean that the fluctuation energies are all the same, when expressed in terms of the mode number $n$, because for each string polarization $ij$ the map $x \leftrightarrow n$ is different and given by (\ref{map}).

Formula (\ref{Om}) for the fluctuation energies has a nice physical interpretation which can indeed be used as an alternative (and simpler) derivation of this expression. Namely, we know \cite{Kazakov:2004qf, Gromov:2007aq} that the quantum fluctuations located at $x$ in the algebraic curve carry energy and momentum\footnote{An extra excitation is still described by formulas (\ref{momQ},\ref{QQ},\ref{epsQ}) with $Q=1$. Quantum fluctuations in the scaling limit have momentum $q\sim \mathcal{O}(1/\sqrt{\lambda})$ so that $X_q^\pm$ are then given by $X^\pm_q=x\pm \frac{4\pi}{\sqrt{\lambda}} \frac{x^2}{x^2-1}$, where $x$ is the position of the quantum fluctuation in the algebraic curve (or the position of the Bethe roots in the AdS/CFT Bethe ansatz \cite{Beisert:2005fw}). Thus in this limit (\ref{fluctuation}) follows from  (\ref{momQ},\ref{epsQ}).}
\beq
\epsilon_{fluct}(x)=\frac{2}{x_n^2-1} \, , \qquad  p_{fluct}(x)=\frac{x}{x_n^2-1}  \,.
\la{fluctuation}
\eeq
When we add a fluctuation the energy is not simply given by $\epsilon_{fluct}(x)$ because the fluctuation will backreact and slightly modify the classical background. The correct energy shift $\Omega$ is then given by the energy of the fluctuation plus the backreaction of the the giant magnon, 
\beq
\Omega(x)=  \epsilon_{fluct}(x)+\({\epsilon_\infty^Q}(p)\)' \delta p \,, \nn
\eeq
where $\delta p$ is the shift in the momentum of the magnon due to the presence of the fluctuation, and  by momentum conservation it equals $-\delta p_{fluct}(x)$. Since
\beq
\({\epsilon_\infty^Q}(p)\)' =2\,\frac{X^+ +X^-}{X^+ X^-+1} \,, \nn
\eeq
formula (\ref{Om}) follows.


\subsection{GM one-loop shift} \la{GM1loop}

The ground state energy around a classical solution, known as the one-loop energy shift, is given by the graded sum over one half of each fluctuation energy:
\beq
\delta \epsilon_{1-loop}=\frac{1}{2} \sum_{n\in \mathbb{Z}}\sum_{(ij)} (-1)^{F_{ij}} \Omega_n^{ij} \,. \nn
\eeq
The sum $(ij)$ extends over all polarizations as listed in (\ref{Stacks}).
It is convenient \cite{SchaferNameki:2006gk, Gromov:2007cd} to rewrite this sum over $n$ as an integral
\beq
\delta \epsilon_{1-loop}=-\sum_{(ij)}(-1)^{F_{ij}} \cw\oint_{\mathbb R}\frac{dn}{4i}\cot\(\pi
n\)\Omega(x_n^{ij}) \,, \nn  
\eeq
where the contour encircles the real axis. Next, for each polarization $(ij)$ we change the variables to $x$ via (\ref{map}) so that the integration over $n$ maps to contours which encircle the fluctuation positions $x_n^{ij}$ located outside the unit circle $\mathbb U$. For a generic polarization $(ij)$, we can deform this contour\footnote{Notice that the orientation of the contour flips in the process.} in the $x$ plane and obtain an integral over the unit circle plus an integral over the eventual cuts of the classical solution, in this case the log cut in (\ref{ps}). Up to exponentially suppressed terms, which we are now focussing on, these two contributions were universally analyzed in \cite{Gromov:2007cd} and \cite{Gromov:2007ky} respectively. The possible contribution from the integral over the log cut is given by the log branch points $x=X^\pm$, which become pole terms when plugged into the integral in the $x$ plane (see (\ref{CurveResult}) below). It is easy to see that this contribution enters as 
$$  
\mathcal{O}\( e^{-\frac{2\pi \Delta}{ \sqrt{\lambda} \sin\frac{p}{2}}} \) \,
$$
and is subleading compared with the contribution coming from the integral over the unit circle which we will now consider. We shall however return to this point in the discussions.

We therefore have
\beqa
\nn \delta \epsilon_{1-loop} =-\sum_{(ij)}(-1)^{F_{ij}} \ccw\oint_{\mathbb
U}\frac{dx}{4i}\frac{p_i'-p_j'}{2\pi}\cot\(\frac{p_i-p_j}{2}\)\Omega(x) \,.
\eeqa
On the upper/lower half of the unit circle, $\mathbb U^{\pm}$, the quasimomenta are large and 
\beqa
\la{expan} \cot\(\frac{p_i-p_j}{2}\)=\pm i  (1+ 2  e^{\mp i (p_i-p_j)} +\dots) \,. 
\eeqa
Picking the leading term we get a zero result for the one-loop shift since
\beqa
 \nn \sum_{ij}(-1)^{F_{ij}} (p_i'-p_j')=0 \,,
\eeqa
which is precisely the vanishing result derived in \cite{Minahan:2006bd} and mentioned in the introduction in (\ref{exp}). The leading exponential correction to the one-loop shift - which in this case turns out to be the leading result for this quantity - is obtained by picking the second term of the expansion of the cotangent. After an integration by parts we find\footnote{The extra factor of $2$ appearing comes from keeping the integration over the upper half circle only. }
\beqa\label{CurveResult}
\delta \epsilon_{1-loop}=\cw\oint_{  \mathbb U^+ } \frac{dx}{2\pi i}\d_x\Omega (x)\sum_{(ij)}(-1)^{F_{ij}}e^{-i(p_i-p_j)}  \,,
\eeqa
where $\Omega(x)$ is given by (\ref{Om}). From (\ref{ps}) using that for a single giant magnon $1/X^{\mp} = X^{\pm}$ including the twists given in (\ref{twists}), we obtain
\beqa
\sum_{(ij)} (-1)^{F_{ij}}e^{-i(p_i-p_j)}
=
\(2 \,\frac{x-X^-}{x-X^+} \,e^{i p/2}-2\)^2
\exp\(-\frac{i4 \pi \Delta}{\sqrt{\lambda}}\frac{x}{x^2-1}\) \la{sump} \,.
\eeqa
It is clear that this expression is suppressed by $e^{-2\pi \Delta/\sqrt{\lambda}}$ which is indeed leading compared to the effects we have discarded so far.  
The expansion (\ref{qt}) mentioned in the introduction is trivially obtained by expanding
 around the saddle point $x\simeq i$.


\section{L\"uscher F-term}\label{sec:Fterm}

The L\"uscher F-term is given by (\ref{FtermdeltaE}). We will now evaluate this for a single magnon and show that it gives rise to the same one-loop leading exponential correction to all orders in $(\Delta/\sqrt{\lambda})^{-1}$ as in (\ref{CurveResult}).
Physically, the F-term computes the correction to the energy of the giant magnon with momentum $p$ by a virtual particle with  momentum $q$ circulating in the compact worldsheet direction. Due to the supertrace in (\ref{FtermdeltaE}), whose origin we discuss in appendix B,  the only contribution comes from the S-matrix part.

The momenta of both the physical particle, $p$, and the virtual particle, $q^*$, appear in the S-matrix through $X_p^\pm\equiv X^\pm$ and $X_{q^*}^\pm$. As mentioned before,  the former lie in the unit circle with
\beq\label{Xpasymp}
X^{\pm}=e^{\pm ip/2}+\mathcal{O}(1/\sqrt{\lambda}) \,.
\eeq
As for the virtual particle \cite{Janik:2007wt} $q^2+\epsilon_\infty^2(q_*(q))=0$ so that
 for $q\sim \mathcal{O}(1)$ we have $q^*\sim \mathcal{O}(1/\sqrt{\lambda})$ and thus
\beq\label{Xqasymp}
X_{q^*}^\pm = x+\mathcal{O}(1/\sqrt{\lambda})\,.
\eeq
In terms of $x$ we have
\beq
\nn q_* =
\frac{4\pi}{\sqrt{\lambda}}\frac{x}{x^2-1}\;\;,\qquad q=i\frac{x^2+1}{x^2-1}  \,.
\eeq
The contour over the real $q$ axis is mapped to the upper half of the unit circle in the $x$ plane and the saddle point of the F-term, at $q=0$, is mapped to $x=i$, which is the same saddle point we found at the end of the previous section.

Moreover
\beq
1-\frac{\epsilon_\infty'(p)}{\epsilon_\infty'(q^*(q))}=1-\frac{x^2+1}{2x}\frac{X^++X^-}{X^+
X^- +1} \,, \nn
\eeq
so that the F-term (\ref{FtermdeltaE}) becomes
\beq\label{FtermResult}
\delta E_a^F=  \cw\oint_{\mathbb{U}^+} \frac{dx}{2\pi i}\,\d_x\Omega(x) \, e^{-4\pi\frac{iJ}{\sqrt{\lambda}}\frac{x}{x^2-1}}\sum_{b} (-1)^{F_b} S_{ba}^{ba}(q^*(q),p)\,,
\eeq
where $\Omega(x)$ is the same function (\ref{Om}) which appeared in the curve computation! Finally, using the explicit form of the AdS/CFT S-matrix in this setup, which we discuss in Appendix B,
we find that this contributes for the particular kinematics (\ref{Xpasymp},\ref{Xqasymp}) as
\beq\label{FtermStrace}
\sum_{b} (-1)^{F_b}  S_{ba}^{ba}(q^*(q),p) = e^{-4\pi \frac{i(\Delta-J)}{\sqrt{\lambda}}\frac{x}{x^2-1}}
 \(2\, \frac{x-X^-}{x-X^+}\sqrt\frac{X^+}{X^-}-2\)^2 \,.
 \eeq
Thus, combining (\ref{FtermResult}) and (\ref{FtermStrace})  we find precisely the integral (\ref{CurveResult}) appearing in the curve computation. Since our match is at the level of the integrals we have an all order (in $(\Delta/ \sqrt{\lambda})^{-1}$) agreement between both computations!


\section{Discussion and Future Work}

The ground state energy for the giant magnon in finite volume is an exponentially suppressed quantity which appears to have an expansion
\beq
\delta \epsilon_{1-loop} =a_{1,0}\,e^{-\frac{2\pi \Delta}{\sqrt{\lambda}}} + \sum_{n \ge 0,m\ge 1}^{\infty} a_{n,m}  \,e^{-n\frac{2\pi \Delta}{\sqrt{\lambda}}-m\frac{2\pi \Delta}{\sqrt{\lambda}\sin\frac{p}{2}}} \,, \nn
\eeq
where each prefactor $a_{n,m}$ is a nontrivial function of $\Delta=L+\sqrt{\lambda}\sin\frac{p}{2}$ and $p$. We computed the leading contribution, namely $a_{1,0}(\Delta,p)$, and checked that it can be reproduced through an F-term contribution (cf. figure \ref{figFmu}).

Depending on the value of the momentum $p$, the next-to-leading order coefficient is $a_{2,0}$ or $a_{0,1}$. The latter has the typical exponential suppression we find from the $\mu$ term computation \cite{Janik:2007wt} and will surely come from the next to leading correction to the results in \cite{Janik:2007wt}. From the algebraic curve point of view this correction could come from two different places. Firstly, from the discrepancy between the log cut description \cite{Minahan:2006bd} used here and the true giant magnon in finite volume described in \cite{Vicedo:2007rp} which is in fact a two-cut solution (cf. figure \ref{figCurve}). Secondly from the integral of the fluctuation energies over the cuts of the classical solution as discussed in section \ref{GM1loop}. It would be very interesting to match the curve and the $\mu$-term results.

As for the the coefficients $a_{n,0}$, they come from the subleading terms in the expansion (\ref{expan}) of the cotangent, and therefore can be trivially computed from the curve point of view to be
\beqa
a_{n,0}\,e^{-\frac{2\pi n \Delta}{\sqrt{\lambda}}}=\cw\oint_{  \mathbb U^+ } \frac{dx}{2\pi i}\d_x\Omega (x)\sum_{(ij)}(-1)^{F_{ij}}e^{-i n (p_i-p_j)}  \,, \la{an0}
\eeqa
where ${\mathbb  U}^+$ is the upper half of the unit circle, $p_i$ the giant magnon quasimomenta (\ref{ps}) and $\Omega(x)$ the flucuation energies (\ref{Om}).

The remaining coefficients $a_{n,m}$  will most likely come from
virtual processes with $n$ virtual particle loops in addition to $m$ splits into two on-shell particles. It would be extremely instructive to generalize the $\mu$- and F-term results of L\"uscher to allow for such processes. The curve computation might serve as an important guide since in principle these exponential corrections can be independently computed in this formalism. It would be interesting to try to find a more general integrable structure behind these contributions, eventually using the already known $a_{n,0}$ as potential guidelines. 

Another possible direction would be to compute the two (worldsheet) loop correction to the dispersion relation $\epsilon_{\infty}(p)$ of the giant magnon. The leading order term will be of the order $\mathcal{O}(e^{-2\pi \Delta/\sqrt{\lambda}})$ and will come from the next-to-leading expansion in $1/\sqrt{\lambda}$ of the F-term obtained in this paper. This computation would provide a two-loop prediction for the dispersion relation of the giant magnon in finite volume which would be interesting to check against a direct two-loop computation in the spirit of  \cite{Roiban:2007jf}.

Yet another interesting next step is to consider more physical string solutions with zero total momentum. The simplest example is to consider two magnons with momenta $p$ and $-p$ (this setup was studied classically in \cite{Minahan:2008re}).
The computations in the sections above can be trivially generalized both in the curve and in the F-term setup and yield for the leading contribution
\beqa\label{TwoMagExp}
\delta \epsilon^{2\, magnons}_{1-loop}=
\frac{128 \, \sin^2 \frac{p}{2}}{\pi i} \cw\oint_{\mathbb U^+}  \frac{dx\,x^3 \exp\(-\frac{i4 \pi \Delta}{\sqrt{\lambda}}\frac{x}{x^2-1}\)  }{(x^2-1)^2\(x^2-2 i x \sin\frac{p}{2}-1\)^2}   \,, 
\eeqa
for the one-loop shift of the two magnon system in finite volume. Here  $\Delta-L=2 \frac{\sqrt{\lambda}}{\pi}\sin\frac{p}{2}$. As an expansion in large $\Delta/\sqrt{\lambda}$ we find therefore the leading terms
\beqa
\delta \epsilon^{2\, magnons}_{1-loop}=-\frac{8\sin^2\frac{p}{2}  }{\pi  \left(1-\sin
   \left(\frac{p}{2}\right)\right)^2}\frac{e^{-\frac{2 \pi \Delta}{\sqrt{\lambda}} }}{ \(\frac{\Delta}{\sqrt{\lambda}}\)^{1/2}   }
   \left[1
   +\frac{\left(11-3 \sin
   \left(\frac{p}{2}\right)\right)
   }{16 \pi \left(1-\sin
   \left(\frac{p}{2}\right)\right)}\frac{1}{\frac{\Delta}{\sqrt{\lambda}}} +O\left(\frac{1}{\(\frac{\Delta}{\sqrt{\lambda}}\)^2
   }\right)\right] \,. \nn
   \eeqa

Finally it would be interesting to try to extend our computations to the most general setup possible including, not only the more general dyonic giant magnons, but also generic classical solutions. For solutions moving in $S^5$, when $L$ is much larger than all the other filling fractions the leading exponential corrections will behave like
\beq
\mathcal{O}\(e^{-\frac{2\pi \Delta}{\sqrt{\lambda}}}\) \,, \nn
\eeq
since the sum over frequencies, when transformed into an integral in the $x$ plane will be dominated by a saddle point at $x\simeq i$ where the quasimomenta are, to leading order,
\beq
p_i\simeq \pm \frac{2\pi \Delta }{\sqrt{\lambda}} \frac{x}{x^2-1} \,. \nn
\eeq
This is in agreement with the findings in \cite{SchaferNameki:2006gk}.
We shall address these issues in a forthcoming publication \cite{GSV2} where we study the F-term vs. curve approach for generic classical solutions.


\subsection*{Acknowledgements}

We would like to thank T.~Dimofte, A.~ Mikhailov, J.~Minahan, R.~Janik, V.~Kazakov, and K.~Zarembo for many
useful discussions. The work of N.G. was partially supported by French
Government PhD fellowship, by RSGSS-1124.2003.2 and by RFFI project
grant 06-02-16786. 
NG was also partly supported by ANR grant INT-AdS/CFT (contract ANR36ADSCSTZ).
S.S.-N. is supported by a Caltech John A. McCone postdoctoral fellowship.
P.~V. is funded by the Funda\c{c}\~ao para a Ci\^encia e Tecnologia
fellowship {SFRH/BD/17959/2004/0WA9}.
We would like to thank the Isaac Newton Institute in Cambridge for an inspiring atmosphere during the SIS workshop.

\setcounter{section}{0}
\setcounter{subsection}{0}


\appendix{Details for algebraic curve computation}


To find the fluctuation energies we perturb the quasimomenta (\ref{ps}) so that $p_i(x)\to p_i(x)+\delta p_i(x)$. The perturbation are fixed by some simple analytical properties of $\delta p_i(x)$. We add $N_n^{ij}$ fluctuations with mode number $n$ and polarization $(ij)$. This means $\delta p_i(x)$ must have poles at position
\beq
p_i(x_n^{ij})-p_j(x_n^{ij})=2\pi n \,, \nn
\eeq
with residue
\beq
\delta p_i(x)\sim \eta_{i} \,N_n^{ij} \,\frac{\alpha(x_n^{ij})}{x-x_n^{ij}} \,, \la{an1}
\eeq
where the signs of the residues are
\beq
\eta_{\h 1}=\eta_{\h 2}=\eta_{\t 3}=\eta_{\t 4}=-\eta_{\h 3}=-\eta_{\h 4}=-\eta_{\t 1}=-\eta_{\t 1}=1 \,, \nn
\eeq
and
\beq
\alpha(x)\equiv \frac{4\pi }{\sqrt{\lambda}}\frac{x^2}{x^2-1}\,. \nn
\eeq
Analogously to the original quasimomenta, the fluctuation $\delta p_i(x)$ can have poles at $x=\pm 1$ but those must be synchronized
\beq
\{\delta p_{\h 1},\delta p_{\h 2}, \delta p_{\h 3},\delta p_{\h 4},\delta  p_{\t 1},  \delta p_{\t 2},  \delta p_{\t 3},  \delta p_{\t 4}\}\simeq
\frac{\{  \alpha_\pm,  \alpha_\pm,  \beta_\pm,  \beta_\pm|  \alpha_\pm,  \alpha_\pm,  \beta_\pm,  \beta_\pm\}}{x\pm 1} \label{an2} \,.
\eeq
due to the Virasoro constraints. There is also an $x \to 1/x$ symmetry inherited from the coset structure of the connection, which for the fluctuations imposes
\beq
\delta p_{\h 1,\h 4}(x)=-\delta p_{\h 2,\h 3}(1/x) \,\, , \,\, \delta p_{\t 1,\t 4}(x)=-\delta p_{\t 2,\t 3}(1/x) \la{an3} \,.
\eeq
The large $x$ asymptotics of the quasimomenta are fixed by the global charges of the theory and are
\beqa
 \(\bea{c}
\delta p_{\h 1}\\
\delta p_{\h 2}\\
\delta p_{\h 3}\\
\delta p_{\h 4} \\ \hline
\delta p_{\t 1}\\
\delta p_{\t 2}\\
\delta p_{\t 3}\\
\delta p_{\t 4}\\
\eea\) \simeq \frac{1}{x}  \frac{4\pi  }{\sqrt{\lambda} }\!
\(\bea{rrl}
+\delta \Delta/2&+N_{\h 1\h 4}+N_{\h 1 \h 3}&+N_{\h 1\t 3}+N_{\h1\t 4}\\
+\delta \Delta/2&+N_{\h 2\h 3}+N_{\h 2\h 4}&+N_{\h2\t 4}+N_{\h2\t 3} \\
-\delta \Delta/2&-N_{\h 2\h 3}-N_{\h 1 \h 3}&-N_{\t 1\h3}-N_{\t 2\h3} \\
-\delta \Delta/2&-N_{\h 1\h 4}-N_{\h 2\h 4}&-N_{\t 2\h4}-N_{\t 1\h4} \\ \hline
&- N_{\t1 \t4}- N_{\t 1\t 3} &-N_{\t 1\h3}-N_{\t 1\h4} \\
&- N_{\t 2 \t3}- N_{ \t 2 \t4}                         &-N_{\t 2\h4}-N_{\t 2\h3}\\
&+ N_{ \t2 \t3}+ N_{ \t1 \t3}&+N_{\h1\t 3}+N_{\h2\t 3}\\
&+ N_{ \t1 \t4}+ N_{ \t 2 \t4}             &+N_{\h2\t 4}+N_{\h1\t 4} \la{an4}
\eea\) \,.
\eeqa
Finally the fluctuations will backreact onto the original solution thus shifting the log cut. Therefore close to $X^\pm$ the quasimomenta will behave like
\beq
\delta p_{\t2,\t 3} (x) \sim  \partial \log(x-X^\pm) \sim \frac{1}{x-X^\pm} \,\, ,
\qquad
\delta p_{\t1,\t 4} (x) \sim    \frac{1}{x-1/X^\pm} \,. \la{an5}
\eeq
This is the only particularity of the GM when compared to the solutions studied in \cite{Gromov:2007aq} where close to the branch points $x^0$ of the classical solution we had $\delta p_i(x) \sim  \partial \sqrt{x-x_0} \sim \frac{1}{ \sqrt{x-x_0}}$.

The analytical properties (\ref{an1},\ref{an2},\ref{an3},\ref{an4},\ref{an5}) are obviously enough to fix the functions $\delta p_i(x)$ completely -- the steps involved mimicking the ones in \cite{Gromov:2007aq} closely. We find
\begin{eqnarray*}
\delta  p_{\t 1}&=&+\frac{A x +B }{x^2-1}-\sum_{n,j=\t 3\t4\h3\h4} \(\frac{N_n^{\t1 j} \alpha(x_n^{\t1j})}{x-x_n^{\t1 j}}-\frac{N_n^{\t2 j} \alpha(x_n^{\t2j})}{1/x-x_n^{\t2 j}}-\frac{N_n^{\t2 j} \alpha(x_n^{\t2 j})}{x_n^{\t2 j}}\)
-\sum_{\beta=\pm }\(\frac{A^\beta}{1/x-X^\beta}+\frac{A^{\beta} }{X^{\beta}}\)
\\
\delta p_{\t 2}&=&+\frac{A x +B}{x^2-1}-   \sum_{n,j=\t 3\t4\h3\h4} \(\frac{N_n^{\t2 j}
\alpha(x_n^{\t2j})}{x-x_n^{\t2 j}}-\frac{N_n^{\t1 j}
\alpha(x_n^{\t1j})}{1/x-x_n^{\t1 j}}-\frac{N_n^{\t1 j} \alpha(x_n^{\t1 j})}{x_n^{\t1 j}}\)+\sum_{\beta=\pm }\frac{A^\beta}{x-X^\beta} \\
\delta  p_{\t 3}&=&-\frac{C x +D}{x^2-1}+\sum_{n,j=\t 1\t2\h1\h2} \(\frac{N_n^{\t3 j} \alpha(x_n^{\t3j})}{x-x_n^{\t3 j}}-\frac{N_n^{\t4 j} \alpha(x_n^{\t4j})}{1/x-x_n^{\t4 j}}-\frac{N_n^{\t4 j} \alpha(x_n^{\t4 j})}{x_n^{\t4 j}}\) -\sum_{\beta=\pm }\frac{A^\beta}{x-X^\beta} \\
\delta  p_{\t 4}&=&-\frac{C x +D }{x^2-1}+\sum_{n,j=\t 1\t2\h1\h2} \(\frac{N_n^{\t4 j} \alpha(x_n^{\t4 j})}{x-x_n^{\t4 j}}-\frac{N_n^{\t3 j} \alpha(x_n^{\t3 j})}{1/x-x_n^{\t3 j}}-\frac{N_n^{\t3 j} \alpha(x_n^{\t3 j})}{x_n^{\t3 j}}\) +\sum_{\beta=\pm }\(\frac{A^\beta}{1/x-X^\beta}+\frac{A^{\beta} }{X^{\beta}}\)
 \\
\delta p_{\h1}&=& +\frac{A x +B }{x^2-1}+  \sum_{n,j=\h 3\h3\t3 \t4} \(\frac{N_n^{\h1 j} \alpha(x_n^{\h1j})}{x-x_n^{\h1 j}}-\frac{N_n^{\h2 j} \alpha(x_n^{\h2j})}{1/x-x_n^{\h2 j}}-\frac{N_n^{\h2 j} \alpha(x_n^{\h2j})}{x_n^{\h2 j}}\)  \\
\delta  p_{\h 2}&=&+\frac{A x +B}{x^2-1} + \sum_{n, j=\h 3\h4\t3\t4} \(\frac{N_n^{\h2 j} \alpha(x_n^{\h2j})}{x-x_n^{\h2 j}}-\frac{N_n^{\h1 j} \alpha(x_n^{\h1j})}{1/x-x_n^{\h1 j}}-\frac{N_n^{\h1 j} \alpha(x_n^{\h1j})}{x_n^{\h1 j}}\)\\
\delta p_{\h 3}&=&-\frac{C x +D}{x^2-1}-  \sum_{n,j=\h 1\h2\t1\t2} \(\frac{N_n^{\h3 j} \alpha(x_n^{\h3 j})}{x-x_n^{\h3 j}}-\frac{N_n^{\h4 j} \alpha(x_n^{\h4j})}{1/x-x_n^{\h4 j}}-\frac{N_n^{\h4 j} \alpha(x_n^{\h4j})}{x_n^{\h4 j}}\)  \\
\delta p_{\h 4}&=&-\frac{C x +D}{x^2-1}-   \sum_{n,j=\h 1\h2\t1\t2} \(\frac{N_n^{\h4 j} \alpha(x_n^{\h4j})}{x-x_n^{\h4 j}}-\frac{N_n^{\h3 j} \alpha(x_n^{\h3j})}{1/x-x_n^{\h3 j}}-\frac{N_n^{\h3 j} \alpha(x_n^{\h3 j})}{x_n^{\h3  j}}\)
\end{eqnarray*}
where the $x\to 1/x$ symmetry fixes
\beqa
&&B=-\sum_{\beta=\pm}\frac{A^\beta}{X^\beta}+\sum_{i=\t1\t2,j=\h3\h4\t3\t4} N_n^{ij} \frac{ \alpha(x_n^{ij}) }{x_n^{ij}} =- \sum_{i=\h1\h2,j=\t3\t4\h3\h 4} N_n^{ij} \frac{ \alpha(x_n^{ij}) }{x_n^{ij}} \nn\\
&&D=-\sum_{\beta=\pm}\frac{A^\beta}{X^\beta}+\sum_{i=\t3\t4,j=\h1\h2\t1\t2} N_n^{ij} \frac{ \alpha(x_n^{ij}) }{x_n^{ij}} =- \sum_{i=\h3\h4,j=\t1\t2\h1\h 2} N_n^{ij} \frac{ \alpha(x_n^{ij}) }{x_n^{ij}} \nn
\eeqa
while the large $x$ asymptotics yields
\beqa
\nn &&A=\frac{2\pi \delta\Delta }{\sqrt{\lambda}}- \sum_{{}^{i=\h1\h2}_{j=\t3\t4\h3\h4}} N_n^{ij} \frac{ \alpha(x_n^{ij}) }{(x_n^{ij})^2} =-\sum_{\beta=\pm} A^\beta+ \sum_{{}^{i=\t1\t2}_{j=\t3\t4\h3\h4}} N_n^{ij} \frac{ \alpha(x_n^{ij}) }{(x_n^{ij})^2} =-\sum_{\beta=\pm} \frac{A^\beta}{(X^\beta)^2}+ \sum_{{}^{i=\t1\t2}_{j=\t3\t4\h3\h4}} N_n^{ij} \frac{ \alpha(x_n^{ij}) }{(x_n^{ij})^2}  \\
\nn &&C=\frac{2\pi \delta\Delta }{\sqrt{\lambda}} - \sum_{{}^{i=\h3\h4}_{j=\t1\t2\h1\h2  }} N_n^{ij} \frac{ \alpha(x_n^{ij}) }{(x_n^{ij})^2} =-\sum_{\beta=\pm} A^\beta+ \sum_{{}^{i=\t3\t4}_{j=\t1\t2\h1\h2}} N_n^{ij} \frac{ \alpha(x_n^{ij}) }{(x_n^{ij})^2}=-\sum_{\beta=\pm} \frac{A^\beta}{(X^\beta)^2}+ \sum_{{}^{i=\t3\t4}_{j=\t1\t2\h1\h2}} N_n^{ij} \frac{ \alpha(x_n^{ij}) }{(x_n^{ij})^2}
\eeqa
These conditions completely fix $A,B,C,D,A^\pm, \delta\Delta$ and require moreover $\sum_{n,ij}n N_n^{ij}=0$ which is nothing but the string level matching condition. For $\delta\Delta$ the result is given by (\ref{deltaDelta}) with $\Omega(x)$ as in (\ref{Om}).


\appendix{Details for F-term computation}\label{sec:Smatrix}

We now specify all the terms appearing in the F-term integral (\ref{FtermdeltaE}).
For our purposes it is enough to consider the AFS-part \cite{Arutyunov:2004vx} of the S-matrix \cite{Beisert:2005tm, Arutyunov:2006yd, Beisert:2006ib}. The one-loop Hernandez-Lopez correction would contribute one order higher in $1/\sqrt{\lambda}$.

We should now also comment on the $(-1)^F$ insertion in (\ref{FtermdeltaE}).
In theories with bosonic and fermionic degrees of freedom, the self-energy corrections, which are computed by the L\"uscher formulas, enter with different signs depending on the statistics of the particles in the loop. This arises solely through the fermion contractions and was also observed in applications of the L\"uscher formula to chiral perturbation theory in \cite{Koma:2004wz} -- see e.g. formula (10) in the first reference in \cite{Koma:2004wz}.
In the self-energy diagrams $I_{abc}$, $J_{abc}$ and $K_{ab}$ of \cite{Luscher:1985dn, Janik:2007wt}, 
we have external lines $a$ given by the giant magnon, so that the statistics of the internal contributions can be summarized by $(-1)^{F_b}$.
This results in particular in the insertion of $(-1)^{F_b}$ in (\ref{FtermdeltaE})
and the contribution from the S-matrix is then\footnote{Note the different sign of the fermion term $-2 a_6$ compared to \cite{Janik:2007wt, Hatsuda:2008gd}, which is due to the supertrace.
Note, that for the $\mu$-term computation of \cite{Janik:2007wt, Hatsuda:2008gd} this sign was irrelevant as in their limit the $a_6$ term did not contribute. Also, the fact that the supertrace removes the $-1$ term in $\sum_{b} (-1)^{F_b}(S-1)$ (for supersymmetric theories) was neglected there, which however was again irrelevant for the purpose of the evaluation of residues that is necessary for the $\mu$-term. However all these points are crucial for the F-term computation in this paper.}:
\beq\label{Strace}
\sum_{b} (-1)^{F_b} S_{ba}^{ba}(q^*,p) =(2a_1+a_2 - 2 a_6)^2 \frac{x_{q^*}^--x_p^+}{x_{q^*}^+-x_p^-} \frac{1-\frac{1}{x_{q^*}^+x_p^-}}{1-\frac{1}{x_{q^*}^-x_p^+}}\sigma^2(x_{q^*},x_p) \,,
\eeq
where
\begin{eqnarray*}
a_1 &=&
\frac{x_p^--x_{q^*}^+}{x_p^+-x_{q^*}^-}\sqrt{\frac{x_p^+}{x_q^-}}\sqrt{\frac{x_{q^*}^-}{x_{q^*}^+}}\,\,,\qquad
 a_6=\frac{x_{q^*}^+-x_{p}^+}{x_{q^*}^--x_{p}^+}\sqrt{\frac{x_{q^*}^-}{x_{q^*}^+}}\cr
a_2&=&
\frac{(x_{q^*}^-- x_{q^*}^+)(x_p^--x_p^+)(x_p^-+x_{q^*}^+)}{(x_{q^*}^--x_p^+)(x_{q^*}^-x_p^--x_{q^*}^+x_p^+)}\sqrt{\frac{x_p^+}{x_q^-}}\sqrt{\frac{x_{q^*}^-}{x_{q^*}^+}}\cr
\sigma^2(x_{q^*},x_p) &=&
\left(\frac{1-\frac{1}{x_p^- x_{q^*}^+}
   }{ 1-\frac{1}{x_{q^*}^-x_p^+} }\right)^{-2}\left(
   \frac{ 1-\frac{1}{x_p^- x_{q^*}^+} }{ 1-\frac{1}{x_{q^*}^- x_p^-} }
   \frac{ 1-\frac{1}{x_{q^*}^-
   x_p^+} }{ 1-\frac{1}{x_{q^*}^+
   x_p^+} }
   \right)^{-i g
   \left(-x_{q^*}^-+x_p^--x_{q^*}^++x_p^++\frac{1}{x_p^-}-\frac{1}{x_{q^*}^+}+\frac{1}{
 x_p^+}-\frac{1}{x_{q^*}^-}\right)} \,.\nn
\end{eqnarray*}
In the scaling limit (\ref{Xqasymp}) we find
\beq
\nn a_2\simeq \mathcal{O}(1/\sqrt{\lambda}) \,,\qquad
a_6\simeq 1 +  \mathcal{O}(1/\sqrt{\lambda}) \,, \quad
a_1=\frac{x-X^-}{x-X^+}\sqrt{\frac{X^+}{X^-}} +  \mathcal{O}(1/\sqrt{\lambda}) \,,
\eeq
and
\beq
\nn \frac{x_{q^*}^--x_p^+}{x_{q^*}^+-x_p^-} \frac{1-\frac{1}{x_{q^*}^+x_p^-}}{1-\frac{1}{x_{q^*}^-x_p^+}}\sigma^2(x_{q^*},x_p) =
\left(\frac{x-X^+}{x-X^-}\frac{x-1/X^+}{x-1/X^-}\right) e^{-\frac{i4 \pi (\Delta-J-Q)}{\sqrt{\lambda}}\frac{x}{x^2-1}}  +  \mathcal{O}(1/\sqrt{\lambda}) \,.
\eeq
For a simple GM we have $X^\pm=1/X^\mp$ and $\sum_{b} S_{ba}^{ba}(q^*,p)  (-1)^{F_b}$ reduces to the simple result (\ref{FtermStrace}) in the main text.


\appendix{Twists -- orbifolding the GM}
In this appendix we closely follow the approach of \cite{Ideguchi:2004wm,Astolfi:2007uz,Gromov:2007ky}  and most importantly the one of \cite{Beisert:2005he}. The giant magnon solution is a worldsheet soliton whose target space picture is that of a string moving in $S^2 \subset S^5$ with endpoints located on the equator and  moving with the speed of light so that \cite{Hofman:2006xt}
$$
Z(t,x=\pm \infty)=X_5(t,x=\pm \infty)+i X_6(t,x=\pm \infty)=e^{it\pm i p/2 +i\alpha}\,,
$$
where $x$ is the rescaled worldsheet space coordinate which ranges from $-\infty$ to $+\infty$ in the infinite spin limit. Thus, to properly treat the giant magnon we must replace the usual periodic boundary condition by
\beq
Z(t,\sigma=2\pi )=e^{-ip} Z(t,\sigma=0)  \la{bcgm} \,,
\eeq
with all the other coordinates $X \equiv X_1+i X_2$ and $Y=X_3+i X_4$ periodic.
The new boundary conditions (\ref{bcgm}) can be incorporated by a $\mathbb{Z}_{S}$ orbifold with
\beq
\textbf{s}_{\textbf 3}=(-t_1+t_3,t_1-t_2+t_3,t_2)=\frac{S}{2\pi}(0,0,-p) \nn \,,
\eeq
which for the \textit{Higher} Dynkin diagram Bethe equations used in \cite{Beisert:2005he,Beisert:2005fw,Gromov:2007ky} amounts to adding a phase
\beq
e^{\frac{2\pi is_j}{S}}  \,\, , \,\, j=1,\dots,7 \nn \,,
\eeq
to the RHS of the Bethe equations (in \cite{Gromov:2007ky} these twists were written as $e^{i\phi_a-i\phi_b}$, see equation (6.1) there).
The seven twists $s_j$ are given by
\beq
\textbf{s}_{\textbf{Higher}}=(t_1,0,t_1-t_2,2t_2-t_1-t_3,t_3-t_2,0,t_3)=\frac{S}{2\pi}(-p/2 ,0,+p/2 , -p  ,+p/2 ,0,-p/2) \,,\nn
\eeq
which in the language of \cite{Gromov:2007ky} means
\begin{eqnarray*}
(\phi_1,\dots,\phi_8)=(\phi_{\t 1},\phi_{\h 1},\phi_{\h 2},\phi_{\t 2},\phi_{\t 3},\phi_{\h 3},\phi_{\h 4}\phi_{\t 4})=\(-\frac{p}{2},0,0,-\frac{p}{2},+\frac{p}{2},0,0,+\frac{p}{2}\)
\end{eqnarray*}
thus explaining the choice (\ref{twists}) in the main text.


\providecommand{\href}[2]{#2}\begingroup\raggedright\endgroup

%


\end{document}